\documentclass[%
 reprint,
 amsmath,amssymb,
 aps,
]{revtex4-2}

\usepackage{physics}
\usepackage{simpler-wick}
\usepackage{graphicx}
\usepackage{dcolumn}
\usepackage{bm}
\usepackage{hyperref}
\hypersetup{
	colorlinks=true,       
	linkcolor=blue,        
	citecolor=blue,        
	filecolor=magenta,     
	urlcolor=blue,         
}

\begin{document}


\title{Highly accurate \textit{ab initio} thermochemistry via real-space quantum Monte Carlo: Benzene}

\author{Iliya Sabzevari}
\email{iliya.sabzevari@gmail.com}
\affiliation{ Department of Chemistry, The University of Colorado at Boulder, Boulder, CO 80302, USA}

\author{Sandeep Sharma}
\email{sanshar@gmail.com}
\affiliation{ Department of Chemistry, The University of Colorado at Boulder, Boulder, CO 80302, USA}





\begin{abstract}
Real-space quantum Monte Carlo is used to calculate the total atomization energy of benzene. In contrast to orbital-space methods common in quantum chemistry, real-space methods allow results at near the complete-basis-set limit to be immediately obtained, all at a computational cost that scales with the fourth power of system size. We demonstrate this utility using the moderately sized benzene molecule, obtaining energies that agree with experiment and previous results from highly accurate composite methods including HEAT and W2. Due to the low scaling of these algorithms, it opens up the possibility of addressing systems out of reach of quantum chemical methods.
\end{abstract}

\maketitle


\section{\label{sec:intro}Introduction}

Quantum chemistry has long promised a first-principles approach to the computation of thermochemical properties. In this endeavor, the high-accuracy extrapolated \textit{ab initio} thermochemistry (HEAT) protocol has proven itself remarkably successful\cite{tajti2004heat,bomble2006high,harding2008high,thorpe2019high}. Recently, an improved version of this protocol has been introduced allowing the computation of heats of formation with nearly sub-chemical accuracy for many important molecules in chemistry\cite{thorpe2019high}. Roughly, the HEAT protocol assumes the following partition for the total energy of a molecule:
\begin{align}
    E_{\text{HEAT}} &= E_{\text{NBO}} \\ 
    &+ \Delta E_{\text{REL}} + \Delta E_{\text{SO}} \\
    &+ \Delta E_{\text{DBO}} + \Delta E_{\text{ZPE}}.
\end{align}
The first term in the summation is the ground state energy of the molecule according to the \textit{non-relativistic Born-Oppenheimer}\cite{born1927quantentheorie} (NBO) Hamiltonian at near the complete-basis-set (CBS) limit. The following terms are corrections due to effects neglected in the above Hamiltonian. $\Delta E_{\text{REL}}$ includes scalar-relativistic effects from the mass-velocity and the one- and two- electron Darwin terms\cite{cowan1976approximate,martin1983all}. $\Delta E_{\text{SO}}$ corrects for spin-orbit coupling while $\Delta E_{\text{DBO}}$ is the diagonal Born-Oppenheimer (DBO) correction and corrects for deficiencies in the Born-Oppenheimer approximation\cite{born1954dynamical}. Finally, $\Delta E_{\text{ZPE}}$ is a correction that takes into account the zero-point energy.

In the usual HEAT protocol $E_{\text{NBO}}$ is obtained from successively more correlated coupled cluster\cite{bartlett2007coupled} (CC) calculations:
\begin{align}
    E_{\text{NBO}} &= E_{\text{HF}}^\infty + \Delta E_{\text{CCSD(T)}}^\infty \\
    &+ \Delta E_{\text{CCSDT}} + \Delta E_{\text{CCSDT(Q)}} + \Delta E_{\text{core}}.
\end{align}
Terms with infinity in the superscript are values extrapolated to the CBS limit, while those without superscripts are performed with a reasonably chosen basis set to capture the bulk of the correlation effects. The series of corrections after the CBS Hartree-Fock (HF) energy take into account electron correlations up to the level of theory written in the subscripts. Finally, $\Delta E_{core}$ takes into account core electron correlations if the preceding calculations are performed with a frozen core of electrons.

Usually in quantum chemistry, theories are formed in the space of orbitals\cite{pople1970molecular}. This includes the coupled cluster and Hartree-Fock models mentioned above. In this study, we demonstrate that the NBO Hamiltonian energy of moderately sized molecules can be accurately obtained using real-space quantum Monte Carlo (QMC) with compact trial wave-functions that contain only $O(N^2)$ number of parameters ($N$ being a measure of system size). These algorithms directly calculate energies at the CBS limit with a modest $O(N^4)$ computational cost\cite{corrsys,realspace,foulkes2001quantum}. Both the high accuracy and low scaling of the method presented here suggest that it can be used to calculate accurate thermochemistry for larger molecules. 

The rest of this study is organized as follows. In section \ref{sec:theory}, the theory outlining real-space QMC is discussed. In section \ref{sec:details}, computational methods utilized in the study are noted. In section \ref{sec:results} we present results for the atomization energy of benzene. We end with our final remarks in section \ref{sec:con}, the conclusion and \ref{sec:ack}, acknowledgments.

\section{\label{sec:theory}Theory}

In this section we will discuss the principal theories involved in quantum Monte Carlo. We will begin with the wave-functions used, then a brief discussion on variational Monte Carlo (VMC) and diffusion Monte Carlo (DMC)\cite{corrsys,realspace,foulkes2001quantum}.

\subsection{wave-functions}
The wave-functions discussed will be Jastrow mean-field wave-functions of the form
\begin{align}
    \ket{\Psi} = \hat{J} \ket{\Phi}.
\end{align}
The operator $\hat{J}$ is the Jastrow factor that explicitly correlates electrons and $\ket{\Phi}$ is the mean-field reference wave-function that satisfies fermionic parity. We will use a factor that includes one- and two-body terms as well as geminal style factors,
\begin{align}
    \hat{J} &= \hat{J}_1 \hat{J}_2 \hat{J}_g.
\end{align}
The one- and two- body factors are functions of inter-particle distance:
\begin{align}
    \hat{J}_1 &= \exp(\sum_{i,I}f^{Z_I}(\hat{r}_{iI})), \\
    \hat{J}_2 &= \exp(\sum_{i<j}g_{\sigma_i,\sigma_j}(\hat{r}_{ij})).\\
\end{align}
Where $\hat{r}_{iI}$ and $\hat{r}_{ij}$ are distance operators between electron $i$ and nucleus $I$ and between electrons $i$ and $j$. The functions $f$ and $g$ are rational functions of the form introduced by Boys  and  Handy\cite{boys1969determination,boys1969calculation,schmidt1990correlated}. 
The last part of the Jastrow factor will be the geminal style factors\cite{casula2003geminal,casula2004correlated,sorella2007weak,casula2004correlated,van2017suppressing}:
\begin{align}
    \hat{J}_g &= \exp(\sum_{i,j}^{n_{\text{elec}}} \sum_{a,b}^{n_{\text{bas}}} J_{ab}^{\sigma_i\sigma_j} \chi_a (\hat{r}_i) \chi_b (\hat{r}_j)
    + \sum_{i}^{n_{\text{elec}}} \sum_{a}^{n_{\text{bas}}} J_{a}^{\sigma_i} \chi_a (\hat{r}_i) 
    ).
\end{align}
Where $\{ \chi_a(r) \}$ are a set of basis functions from which one- and two- electron expansions are constructed with the variational parameters $\{J_{ab}^{\sigma\sigma'}, J_{a}^{\sigma}\}$.
    
For the reference wave-function $\ket{\Phi}$, we will discuss three forms. The first will be the 
restricted Hartree-Fock (RHF) wave-function\cite{szabo2012modern}, the second will be the $\hat{S}_z$ projected generalized Hartree-Fock ($\hat{S}_z$ GHF) wave-function\cite{jimenez2012projected,valatin1961generalized,bach1994generalized,mahajan2019symmetry}, and the last will be the anti-symmetrized geminal power (AGP) wave-function\cite{coleman1965structure,bardeen1957theory,genovese2020general,mahajan2019symmetry}. All parameters in the wave-functions, including those in the Jastrow factors and reference, are optimized within VMC.

\subsection{variational Monte Carlo}

To evaluate the expectation value of operators in the real-space Monte Carlo framework, we will use variational Monte Carlo (VMC). To this end, one first notes that the expectation value of the operator $\hat{O}$, with respect to any wave function $\Psi$, can be evaluated as the integral, 
\begin{align}
    \expval{\hat{O}}_\Psi = \int dR \rho (R) O_l(R),\\
    \rho(R) = \frac{ \norm{\bra{R}\ket{\Psi}}^2 }{ \bra{\Psi}\ket{\Psi} } \\
    O_l(R) = \frac{ \bra{R}\hat{O}\ket{\Psi} }{ \bra{R}\ket{\Psi} }.
\end{align}
One can then evaluate the above integral via Markov Chain Monte Carlo sampling of the distribution $\rho(R)$. We use the accelerated Metropolis-Hastings algorithm introduced in reference\cite{umrigar1993accelerated} to do this.

The wave-function parameters are optimized using a combination of the linear method\cite{toulouse2008full,sabzevari2020accelerated} (LM) and the stochastic reconfiguration\cite{sorella2001generalized,sorella2007weak,neuscamman2012optimizing} (SR) algorithm. In these calculations we use three steps to optimize the wave-function. In step one, we obtain the reference state using a deterministic algorithm e.g. the RHF state is simply obtained by running a self-consistent field calculation. In the second step, we fix the reference and optimize the Jastrow parameters using the LM, which typically converges in less than 10 iterations. In the final step, all parameters in the wave-function, including those in the Jastrow and the reference, are optimized using the SR algorithm.

\subsection{diffusion Monte Carlo}

In this study we will also employ diffusion Monte Carlo (DMC). This algorithm is part of a class of projector Monte Carlo algorithms that, as the name suggests, involves projecting out the ground state of a system. This is done by noting the formal solution to the imaginary time Schr\"{o}dinger equation in real-space is,
\begin{align}
    \Psi(R',\tau+t) = \int G(R',R,\tau) \Psi(R,t) dR, \\
    G(R',R,\tau) = \bra{R'} \exp(-\tau \hat{H}) \ket{R}.
\end{align}
The function $G(R',R,\tau)$ is the so-called propagator or Green's function and it is used to drive the dynamics of a stochastic simulation. Naive use of this form of the propagator results in a simulation with an exponentially growing signal-to-noise ratio\cite{kalos1974helium}. To remedy this, one employs the \textit{fixed-node} approximation, in which one performs importance sampling using a guiding wave function\cite{anderson1976quantum,reynolds1982fixed}. The resulting importance sampled Green's function is,
\begin{align}
    \bar{G}(R',R,\tau) = \frac{\Psi_G (R')}{\Psi_G (R)} \bra{R'} \exp(-\tau \hat{H}) \ket{R}.
\end{align}
This allows the simulation of fermionic systems to be performed for a given error at a low polynomial cost. In this work, we use the form of propagator recently introduced in reference\cite{anderson2021nonlocal}, including the fully summed T-move\cite{casula2006beyond,casula2010size} terms. With this form of the propagator, we have found an imaginary time step of 0.01 Ha$^{-1}$ to be sufficient to obtain negligible time step errors for all calculations in this study.

In DMC, it is customary to employ an effective core potential (ECP) approximation to the core electrons of the system under study, analogous to the frozen core (FC) approximation common in quantum chemistry\cite{hammond1987valence}. This greatly reduces the cost of DMC, as Monte Carlo sampling the core electrons quickly becomes the dominant cost in the algorithm. The ECP approximation then makes large molecule calculations easier to perform, particularly with atoms of large atomic number. Recently, to obtain results beyond the ECP approximation, \textit{lattice regularized} (LR) DMC has been introduced\cite{casula2006ground,casula2005diffusion}. LR-DMC allows a partitioning of the length scales between core and valence electrons and makes all-electron (AE) calculations of large molecules significantly more tractable\cite{nakano2020speeding}. In Section \ref{sec:results}, we will compare state-of-the-art AE LR-DMC calculations in the literature to core corrected ECP calculations performed for this study.

\section{\label{sec:details}Computational Details}

In this section, details of the computation will be discussed, which includes software used to perform the calculations. For the geminal-style Jastrow factors, we will only use the valence sub-set of the atomic orbitals. Since our principal example is benzene, this includes the Carbon $2s$ and $2p$ orbitals and the hydrogen $1s$ orbitals. To obtain input parameters for the QMC calculations, we use Pyscf\cite{sun2018pyscf,sun2020recent} to perform an initial RHF or GHF calculation. All variational Jastrow parameters will be set to 0 at the start of the optimization. QMC calculations will be performed using custom open-source software developed in our group\cite{github}.

\section{\label{sec:results}Results and Discussion}

In this section, we will discuss numerical results for the total atomization energy of benzene. The first part of the discussion will be on obtaining the energy of the non-relativistic Born-Oppenheimer (NBO) Hamiltonian in the Monte Carlo framework. The second part will be focused on a comparison to other values in the literature from theory and experiment.

\subsection{$E_{NBO}$}

We first present values for $E_{NBO}$ from calculations performed for this study. The core electrons will be approximated using correlation-consistent ECPs (ccECPs) for the reasons mentioned in Section \ref{sec:theory}, in conjunction with the respective cc-pVDZ basis sets\cite{bennett2017new,annaberdiyev2018new}. These values will be corrected for core correlation effects outside of the ECP approximation in a manner analogous to the HEAT protocol:
\begin{align}
    E_{\text{NBO}} &= E_{\text{DMC}}(ecp) + \Delta E_{\text{core}},
\end{align}
where the correction in this case is
\begin{align}
    \Delta E_{\text{core}} &= (E(ae) - E(ecp))_{\text{UCCSD(T)}}.
\end{align}
Total energies and their corrections are shown in Table \ref{tab:dmc}.
\begin{table}[h]
    \centering
    \begin{tabular}{lc|cc|c}
        \hline
        \hline
        wave & $E_{\text{DMC}}(ecp)$ & basis & $\Delta E_{\text{core}}$ & $E_{\text{NBO}}$ \\
        \hline
        DMC-JRHF & 2.180(2) & dz & 0.0149& 2.195(2) \\
         & & tz & 0.0290 & 2.209(2) \\
        \hline
        DMC-J$S_z$GHF & 2.153(1) & dz & 0.0149& 2.168(1) \\
         & & tz & 0.0290 & 2.182(1) \\
        \hline
        \hline
    \end{tabular}
    \caption{Total $E_{\text{NBO}}$ in the right-most column. First column is the guiding wave-function used in DMC with the corresponding computed energy. The middle column is the computed corrections using a cc-pVDZ or cc-pVTZ basis set. Energies are in units of Hartrees.}
    \label{tab:dmc}
\end{table}

Literature values using LR-DMC are also presented. For reasons already discussed, these are particularly noteworthy as all-electron (AE) energies at the complete-basis-set (CBS) limit can be immediately obtained:
\begin{align}
    E_{\text{NBO}} &= E_{\text{LRDMC}}(ae).
\end{align}
We present these energies in Table \ref{tab:lrdmc}.
\begin{table}[h]
    \centering
    \begin{tabular}{l|c}
        \hline
        \hline
        wave & $E_{NBO}$ \\
        \hline
        LRDMC-JRHF\cite{genovese2020general} & 2.2083(7) \\
        LRDMC-JAGP\cite{genovese2020general} & 2.1741(7) \\
        \hline
        \hline
    \end{tabular}
    \caption{$E_{NBO}$ from all-electron LR-DMC calculations using the guiding wave-functions on the left. A basis set with a mixture of cc-pVDZ and cc-pVTZ basis functions was used for the calculations. Energies are in units of Hartrees.}
    \label{tab:lrdmc}
\end{table}
Comparing the two sets of energies, we obtain remarkable agreement of the NBO energy using the UCCSD(T) core correction with the cc-pVTZ basis set. This is no doubt due to a fortuitous cancellation of error in the computed quantities. Our DMC energies obtained using the RHF reference (see Table \ref{tab:dmc}) as the guiding wave-function agree to within a mHa of the LRDMC energies of Sorella et al. (see Table \ref{tab:lrdmc}) with the same reference, which is within the stochastic error of the calculations. The JAGP reference wave-function delivers an atomization energy slightly smaller than the one obtained with our J$S_z$GHF wave function, but there is reasonable agreement nonetheless.

Calculations that have been presented are summarized in Table \ref{tab:sum} along with other values for the NBO energy from other theories.
\begin{table}[h]
    \centering
    \begin{tabular}{l|l}
        \hline
        \hline
        theory & $E_{\text{NBO}}$ \\
        \hline
        DMC-JRHF/UCCSD(T) & 2.209(2) \\
        DMC-J$S_z$GHF/UCCSD(T) & 2.182(1) \\
        LRDMC-JRHF\cite{genovese2020general} & 2.2083(7) \\
        LRDMC-JAGP\cite{genovese2020general} & 2.1748(7) \\
        HEAT'345-(Q)\cite{harding2011towards} & 2.1820 \\
        W2\cite{parthiban2001fully} & 2.1835 \\
        \hline
        \hline
    \end{tabular}
    \caption{$E_{NBO}$ for various orbital-space and real-space electronic structure methods. Energies are in units of Hartrees.}
    \label{tab:sum}
\end{table}
The HEAT'345-(Q) result is highly accurate using large-scale coupled cluster calculations performed with all 42 electrons correlated and more than 1500 basis functions, as well as including contributions from quadruple excitations\cite{harding2011towards}. The Weizmann-2 (W2) calculation agrees within chemical accuracy ($\sim 1$ mHa) to the HEAT'345-(Q) result, however the calculations are somewhat smaller in comparison, with the largest calculations involving CCSD/cc-pV5Z and CCSD(T)/cc-pVQZ\cite{parthiban2001fully}. The real-space QMC results using the high quality reference wave-functions give good agreement with both W2 and HEAT'1345-(Q) energies. To make a comparison with experiment one must include the other effects included in the HEAT protocol; this will be the subject of the next subsection.

\subsubsection{total atomization energy}

As mentioned in the introduction, to obtain the total atomization energy (TAE), one needs to include corrections to the NBO energy for scalar-relativistic, spin-orbit, zero-point energy, and diagonal Born-Oppenheimer effects. Obtaining these values in real-space QMC is beyond the scope of this paper, as they are active areas of research\cite{melton2016spin,mccoy2006diffusion}. We will borrow the values obtained from previous thermochemical studies, principally the HEAT'345-(Q) results we discussed partially above\cite{harding2011towards}. The energy corrections are summarized in Table \ref{tab:corrections}. Note that the energy corrections computed for the W2 calculation are practically the same numerical values, except $E_{DBO}$ is taken to be zero.
\begin{table}[]
    \centering
    \begin{tabular}{l|r}
        \hline
        \hline
        $\Delta E_{\text{ZPE}}$ & -0.0990 \\
        $\Delta E_{\text{REL}}$ & -0.0016 \\
        $\Delta E_{\text{SO}}$ & -0.0009 \\
        $\Delta E_{\text{DBO}}$ & 0.0002 \\
        \hline
        \hline
    \end{tabular}
    \caption{Various energy corrections to the non-relativistic Born-Oppenheimer energy of benzene\cite{harding2011towards}. Energies are in units of Hartrees.}
    \label{tab:corrections}
\end{table}
The zero-point vibrational contribution was obtained with vibrational second-order
perturbation theory\cite{mills19723} (VPT2) using force-fields calculated via CCSD(T) with a quadruple-zeta quality basis set. The scalar-relativistic contribution was obtained by computing the corresponding perturbative corrections due to the mass-velocity and the one- and two-electron Darwin terms\cite{cowan1976approximate,martin1983all} with CCSD(T)/aug-cc-pCVTZ. The spin-orbit contribution is also taken into account to first-order and is evaluated with SO-MRCISD (spin-orbit multi-reference configuration interaction singles and doubles\cite{tilson2000parallel}) employing a double-zeta basis\cite{tajti2004heat} and relativistic ECPs\cite{fernandez1985ab}. The DBO correction is the first-order correction to the electronic energy of the nuclear kinetic energy operator, it was calculated with CCSD/aug-ccpCVTZ as described in reference \cite{gauss2006analytic}. We summarize TAE results for the high quality calculations in Table \ref{tab:tae}. There is exceptional agreement between all theoretical values and experimental results from the active thermochemical tables (ATcT).
\begin{table}[]
    \centering
    \begin{tabular}{l|ll}
        \hline
        \hline
        theory & Ha & kJ/mol \\
        \hline
        DMC-J$S_z$GHF/UCCSD(T) & 2.081(1) & 5463(3). \\
        LRDMC-JAGP\cite{genovese2020general} & 2.0735(7) & 5444(2).\\
        HEAT'345-(Q)\cite{harding2011towards} & 2.0807 & 5463.0 \\
        W2\cite{parthiban2001fully} & 2.0822 & 5466.7 \\
        \hline
        ATcT\cite{stevens2010heats} & 2.0811 & 5463.8 \\
        \hline
        \hline
    \end{tabular}
    \caption{Total atomization energy labeled by the principal theory involved. Experimental results from the active thermochemical tables (ATcT) are also presented.}
    \label{tab:tae}
\end{table}

\section{\label{sec:con}Conclusion}

In this article, we have briefly demonstrated that real-space QMC can be used to calculate high accuracy thermochemistry of moderately sized molecules. Quite notably, we obtain results that rival highly correlated coupled cluster theories using basis sets with almost on order of magnitude fewer functions. This highlights the effectiveness real-space QMC produces without the need for costly basis set extrapolations. Further, due to the modest $O(N^4)$ scaling of the calculations with system size, one can do similar calculations on significantly larger systems. 

In the future, we will focus on larger systems that are beyond the reach of high accuracy methods such as HEAT. Future work will also focus on developing the lattice-regularized approach to DMC, which will allow one to perform all-electron calculations at a reasonable cost. 
 Work will also be done to obtain other energy contributions, such as zero-point energy and spin-orbit coupling, in the real-space QMC framework.

\section{\label{sec:ack}Acknowledgments}

The majority of funding for this project was provided by NSF, through the grant CHE-1800584. SS was partly supported through the Sloan research fellowship. IS was partly supported through the Molssi fellowship. All calculations were performed on the Summit cluster at CU Boulder. IS would like to give a special thank-you to John F. Stanton for introducing him to the wonderful world of quantum chemistry.


%

\end{document}